# Progress of the anti-obesity of Berberine

Yue Kong[1], Haokun Yang[2], Rong Nie[1], Xuxiang Zhang[1], Hongtao Zhang[3,]*  & Xin Nian[1,4,]*

**Abstract：**

Obesity is defined as the excessive accumulation or abnormal distribution of body fat. According to data from World Obesity Atlas 2024, the increase in prevalence of obesity has become a major worldwide health problem in adults as well as among children and adolescents. Although an increasing number of drugs have been approved for the treatment of obesity in recent years, many of these drugs have inevitable side effects which have increased the demand for new safe, accessible and effective drugs for obesity and prompt interest in natural products. Berberine (BBR) and its metabolites, known for their multiple pharmacological effects. Recent studies have emphatically highlighted the anti-obesity benefits of BBR and the underlying mechanisms have been gradually elucidated. However, its clinical application is limited by poor oral absorption and low bioavailability. Based on this, this review summarizes current research on the anti-obesity effects of BBR and its metabolites, including advancements in clinical trail results, understanding potential molecular mechanisms and absorption and bioavailability. As a natural compound derived from plants, BBR holds potential as an alternative approach for managing obesity.

## 1. Introduction

As we known, obesity is a global problem. With over 2 billion people being

1  Department of Endocrinology, The First Affiliated Hospital of Kunming Medical University, Kunming, China
2  Kunming Medical University, Kunming, China
3  Hangzhou CytoCan Biotech, Hangzhou, 310029, China.
4  Lead contact
*Corresponding author:Xin Nian and Hongtao Zhang
*Correspondence: nianxinkm@hotmail.com(X.N.), zhanghongtao@cytocan.com (H.T.Z.)

overweight (BMI ≥ 25kg/m²), of these about 600 million people are having obesity (BMI ≥ 30kg/m²) [1]. Of course, if the Asia-Pacific classification of obesity recommended by the World Health Organization (WHO)(overweight: BMI ≥ 23 kg/m², obesity: BMI ≥ 25 kg/m²) was applied, no doubt about that the actual number of obesity will be much higher. Most of these people are living in emerging economies (BRICS countries) and in rapidly developing countries of Asia, Africa, and South America [2]. According to the specific data from World Obesity Atlas 2024, there were over 800 million adults living with obesity worldwide in 2020 and current trends predict a prevalence of over 1.5 billion in 2035 [3]. Above these evidences, it suggest that obesity has become a major public health problem in the global scale.

Obesity is widely recognized as a complex and multifactorial disease. Pathophysiology of obesity primarily involves out of balance in energy homeostasis and metabolic adaptation, hormonal regulation, neural control, inflammation and immune responses, molecular mechanisms (genetic factors and epigenetic modifications), gut microbiome. Based on this, the incidence, prognosis and treatment effect of obesity itself and various complications caused by obesity are showing a bad situation[4][5][6][7][8]. Currently recommended therapies with evidence-based support in obesity mainly are lifestyle intervention [9], pharmacotherapy [10] and bariatric surgery [11][12].

Lifestyle intervention, known as diet, exercise or behavior therapy, is the recommended initial therapy for improving weight and obesity-related comorbidities in patients with obesity. However, most studies have shown that lifestyle intervention has the disadvantages of poor patient compliance, unreasonable intervention measures, poor clinical effect of weight loss and high risk of weight regain [13][14][15]. For many obese patients, anti-obesity drugs remain the first-line treatment. Anti-obesity drugs are approved in patients with a BMI ≥ 27 kg/m² at least one obesity-related comorbidity or in patients with a BMI ≥ 30 kg/m². Currently approved anti-obesity medications include naltrexone/bupropion, orlistat, liraglutide, semaglutide, tirzepatide and phentermine/topiramate. However, the clinical effect and safety issues of current anti-obesity drugs have brought great challenges to the tratment of obesity

[5]. For instance, amphetamines and phentermine may lead to serious cardiovascular events, such as arrhythmias and increased blood pressure. Orlistat, which inhibits fat absorption, can cause gastrointestinal issues like steatorrhea (fatty stools) and constipation. In rare cases, these medications have been linked to sudden death [16]. In addition, aforementions drugs including naltrexone/bupropion, orlistat and phentermine/topiramate also were reported that have kidney injury and induce kidney disease[17][18][19][20][21]. GLP-1RAs, new weight loss therapeutics, are often only used on a short-term basis mainly due to high costs and adverse side effects. According to existing research results, most side effects of GLP-1RAs commonly involve gastrointestinal symptoms, pancreatic safety (pancreatitis and pancreatic cancer), thyroid cancer, gallbladder events and injection-site and allergic reactions [22][23][24][25]. Bariatric surgery alters gastrointestinal anatomy impacting food intake and/or nutrient absorption, which currently targeting patients with BMI ≥ 40kg/m$^2$ or BMI ≥ 35kg/m$^2$ and a comorbidity or failed medical treatment[25][26][27][28]. However, bariatric surgery has limited use due to perceived invasiveness, high cost and postoperative adverse events, including negatively affect bone health, dumping syndrome, VD deficiency, postoperative pain, nausea or vomiting, hypostatic pneumonia, deep venous thrombosis (DVT), pulmonary embolism, stress ulcers [29][30][31][32][33][34][35][36][37].

The inevitable side effects of aforementioned current obesity therapeutics have increased the demand for new treatments and prompt interest in natural products. Coptidis Rhizoma (the rhizome of Coptis chinensis), commonly known as Huang Lian in China, Ouren in Japan and Hwang-Ryun in Korea, is notable for its diverse alkaloid content [38]. From the n-butanol (BuOH) fraction of the methanol (MeOH) extract of Coptidis Rhizoma, several active constituents including BBR(figure 1), epiberberine, coptisine, palmatine and magnoflorine have been isolated, and BBR is the most predominant and attractive component [39]. BBR is also found in various medicinal plants such as Berberis aristata, B. petiolaris, B. aquifolium, B. vulgaris, B. thunbergia and many others [40]. The four major metabolites of BBR are berberrubine

(M1), thalifendine (M2), demethyleneberberine (M3), and jatrorrhizine (M4) [41].

**Figure 1 Chemical Construction of BBR**

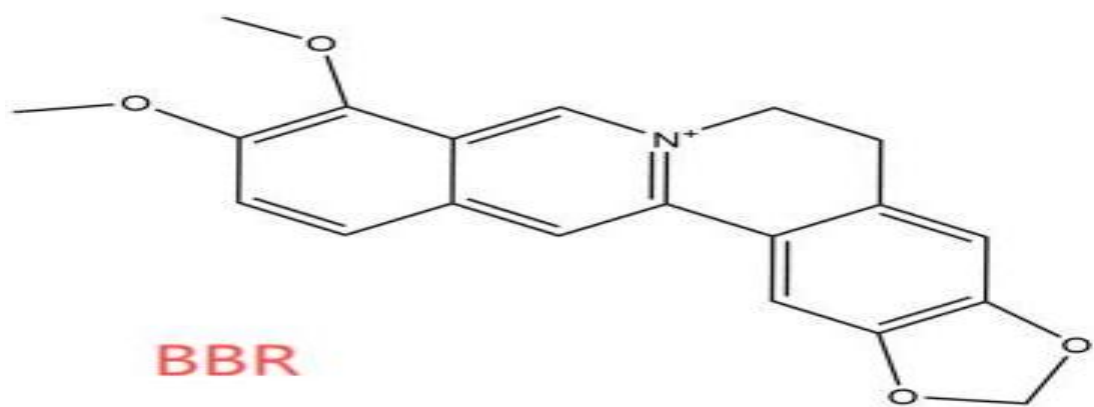

BBR has been used as an over-the-counter drug in China to treat diarrhea caused by bacteria for more than 70 years, shown very good safety profiles [42]. In addition, BBR has other various pharmacological and biological effects, particularly the context of obesity management has received much attention in recent years [38] [42] [43]. A large number of researches have consistently shown that BBR and its derivatives have favourable clinical effects and safety in the treatment of obesity and related complications. From a mechanistic point of view, Poulios et al. summarized the beneficial effects of the most important phytochemicals that are represented by BBR against obesity. Mechanisms of action include inhibition of adipocyte differentiation, browning of the white adipose tissue (WAT), inhibition of enzymes such as lipase and amylase, suppression of inflammation, improvement of the gut microbiota, and downregulation of obesity-inducing genes [44]. In clinical studies, oral administration is the predominant method for BBR delivery, with the primary site for absorption and metabolism being the intestine [45]. However, the absolute bioavailability of BBR has been demonstrated to be less than 1% in rats [46]. The poor absorption is attributed to factors such as low permeability, hepatobiliary re-excretion, P-glycoprotein-mediated efflux, and self-aggregation. Furthermore, the first-pass metabolism in the liver and intestine also contributes to its low oral bioavailability [47]. Enhancing BBR absorption would represent a significant advancement for the pharmaceutical development of BBR.

Based on the above-mentioned, this summary will summarize current researches on the anti-obesity efficacy of BBR and its metabolites, focus on advancements in clinical trail results, molecular mechanisms and absorption and bioavailability methods.

## 2. Clinical trail results about anti-obesity of BBR

Numerous recent preclinical and clinical research have increasingly focused on exploring deeper insights into the anti-obesity effects of BBR.

### 2.1 Effectiveness of BBR against obesity

### 2.1.1 Preclinical research（Table 1）

Xu et al. reported that treatment with BBR (150 mg/kg/day) for 6 weeks significantly reversed the body weight increase in HFD-fed rats, as compared with untreated HFD-fed rats, without interfering with food intake[48]. Comparing with control group, Park et al. verified that the HFD-induced obesity mice received daily intraperitoneal injections of BBR (10 mg/kg) for 3 weeks significantly reduced food intake, body weight, fat contents, serum leptin, and glucose level[49]. In diet-induced obesity in rats, Wang et al. reported the novel discovery that the mesenteric small arterioles are over dilated, NO is overproduced and NA production is decreased in mesenteric perivascular AT (PVAT). All these changes were reversed by BBR (200 mg/kg) through ameliorates vascular dysfunction by regulating the overproduction of NO in PVAT[50]. Yang et al. found that orally administration of BBR (200 mg/kg) decreased serum and liver lipid levels in HFD-induced hyperlipidemic mice. Further analysis revealed that the BBR-enriched gut commensal Blautia producta ameliorates HFD-induced hyperlipidemia, stimulates liver LDLR expression and promoted LDL uptake by HepG2 cells[51]. Sun et al. had a study observed that treatment with the oral administration of BBR (100 mg/kg/d, 200 mg/kg/d) alleviated HFD-induced obesity in mice [52]. Wang et al. showed that BBR obviously reduced body weight in diet-induced obese mice, at both low dose (150 mg/ (kg·d) by gavage) and high dose

(300 mg/ (kg·d)) for 4 weeks, without apparent appetite suppression [53]. Likewise, BBR (100 mg/kg, po) significantly ameliorated insulin resistance, reduced body weight and percent body fat and improved serum biochemical parameters and obesity effectively in HFD-mice [54]. Gupta et al. also verified that BBR at the doses 100, 200, and 500 mg/kg (p.o.) ameliorated the total fat mass in animal models of dexamethasone (Dex) - induced to varying degrees [47].

**Table 1 Preclinical researches**

| Author | Processing Scheme | Result |
|---|---|---|
| Xu et al. | Treatment with an oral supplement of BBR for 6 weeks (150 mg/kg/day) | Treatment with BBR for 6 weeks significantly reversed the body weight increase of HFD-fed rats compared with untreated HFD-fed rats (658.58 ± 54.04 vs. 715.59 ± 46.70 g; $P<0.05$). |
| Park et al. | The mice received daily intraperitoneal injections of BBR (10 mg/kg) for 3 weeks | After BBR treatment, body weight was significantly decreased (on 66th, 69th, 72th, and 75th day: $P< 0.05$), the relative epididymal ($P < 0.05$) and peritoneal fat mass ($P<0.01$) significantly reduced compared to HFD-fed mice. The level of leptin decreased in treated with BBR (30.8 ± 13 ng/ml) compared to HFD-fed mice (39.8 ± 4.5 ng/ml). |
| Wang et al. | Obese rats were treated with BBR (200 mg/kg) by gavage for 2 weeks | The administration of BBR significantly inhibited weight gain ($P< 0.01$), significantly improved blood sugar ($P< 0.05$) and blood lipid abnormalities ($P< 0.01$), and simultaneously corrected inflammation (IL-6, IL-1β and TNF-α, $P < 0.01$). |
| Yang et al. | Administration of BBR via oral gavage (200 mg/kg) for 56 | Treatment with BBR significantly decreased body weight gain ($P < 0.001$) and serum lipids ($P < 0.05$) such as total |

| | days | cholesterol (TC), triglycerides (TG) and low-density lipoprotein cholesterol (LDL-C), and the HFD-induced liver steatosis (P < 0.05). |
|---|---|---|
| Sun et al. | Oral administration of BBR(100mg/kg/d, 200mg/kg/d, 5times/week) | The oral administration of BBR alleviated HFD-induced obesity by a dose-dependent manner in mice that U73122 partially inhibited. BBR upregulated the release of GLP-1, promoted the proliferation of tuft cells and secretion of IL-25 in obesity via the TAS2R signaling pathway. |
| Wang et al. | Treatment with low dose (150 mg/(kg·d)) and high dose (300 mg/(kg·d)) for 4 weeks of continuous gavage | Both low and high doses of BBR effectively reduced the body weight (P < 0.001), reversed insulin resistance and abnormalities in glucose tolerance, reduce serum TG (P < 0.001), TC (P < 0.001), and LDL-C (P < 0.01) levels and elevate serum HDL-C (P < 0.001) level of DIO mice. |
| Du et al. | Treatment with BBR (100 mg/kg, po) | The oral administration of BBR significantly reduced the AUCs of the OGTT (p < 0.01) and the ITT (p < 0.05), the serum biochemical parameters, including FBG, TC, TG and LDL-C (p < 0.01), and volume of lipid droplets (p < 0.05). |
| Gupta et al. | Treatment with BBR at the doses of 100, 200, and 500 mg/kg (p.o.) in mice were treated with Dex (2 | BBR improved glucose homeostasis in diabetic mice by enhancing glucose clearance, increasing glycolysis, elevating glucose uptake, and decreasing gluconeogenesis. Further, Dex treatment increased the total fat |

| | mg/kg, i.m.) for 30 days | mass in mice, which was ameliorated by BBR treatment (p < 0.01). |

### 2.1.2 Clinical research (Table 2)

Hu et al. investigated that BBR supplementation (1.5 g/d) for 12 weeks significantly reduced blood lipid levels in obese subjects, including a 23% reduction in TG and a 12.2% reduction in TC levels [55]. Bandala et al. testified that orally administration of BBR (500 mg/tablet, three times a day before each meal lasted 3 months) significantly decreased (p < 0.05) body weight, BMI, body fat percentage and visceral fat percentage in obese patients [56].

## Table 2 Clinical researches

| Author | Processing Scheme | Result |
| --- | --- | --- |
| Hu et al. | Obese human subjects (Caucasian) were given 500 mg BBR orally three times a day for 12 weeks | BBR treatment produced a mild weight loss (average 5lb/subject) in obese human subjects, significantly reduced blood lipid levels (23% decrease of TG and 12.2% decrease of TC levels) in human subjects, and reduced in the rat experiment (34.7% decrease of TG and 9% decrease of TC level). |
| Bandala et al. | Obese patients were given orally administration of BBR (500 mg/tablet, three times a day before each meal) lasted 3 months | BBR significantly decreased (p < 0.05) body weight, BMI, body fat percentage, visceral fat percentage, and diastolic blood pressure. |

In summary, all the above preclinical and clinical data demonstrated the anti-obesity effect of BBR, and indicated BBR's potential in the treatment of obesity. Besides, Gasmi et al. analyzed the representative studies for the years 1982-2022. Literature analysis shows that the evidence of BBR appears to be sufficient for hands-on use for obesity [57].

**2.2 Safety of BBR against obesity**

Many previous studies have demonstrated that BBR is effective and safe alternatives to synthetic drugs as a natural product. A randomized double-blind placebo-controlled clinical trial, which aimed to evaluate the potential benefit of oral supplementation of BBR PhytosomeTM (2 tablets/day, 550 mg/tablet) on the metabolic profile of subjects with impaired fasting blood glucose (IFG), indicate that the use of berberine had no adverse events [58]. However, some studies have reported adverse reactions mainly including gastrointestinal symptoms of BBR. In Bandala's study, the main reported adverse effects in patients who were treated with BBR were nausea (20%), constipation (16%) and hemorrhoidal bleeding (28%), but these symptoms decreased within the first month of treatment. Clinical studies have shown that long-term application of BBR will bring side effects of constipation.

In response to the above, Quercetin (QR) was found to be the effective ingredient of Amomum Villosum Lour (AVL) in terms of relieving constipation in Cui's study, the use of QR in combination with BBR has an adverse effect-reducing efficacy [59]. To sum up, BBR is a safe medicinal plant ingredient. Additionally, the clinical effects of BBR need to be confirmed in high-quality RCTs.

**3.The potential mechanisms of BBR against obesity**

**3.1 The activity of BBR in suppressing adipocyte differentiation**

Preadipocytes undergo a process of adipogenesis, transforming into differentiated adipocytes. This process is characterized by increased expression of adipogenic transcription factors and adipocyte-specific genes [38]. Inhibiting these transcription factors and adipocyte-specific genes may be a key mechanism in the suppression of adipogenesis.

**3.1.1 Down-regulation of C/EBP-α and PPAR-γ activities and up-regulation of PPAR-δ on 3T3-L1 adipocytes by BBR**

Adipogenic differentiation is a continuous transcriptional process: CCAAT/enhancer-binding protein-β (C/EBP-β) and CCAAT/enhancer-binding protein-δ (C/EBP-δ) are transiently induced by hormonal signals, and subsequently induce directly the expression of peroxisome proliferator-activated receptor γ (PPAR-γ) and CCAAT/enhancer-binding protein-α (C/EBP-α). This activation leads to the expression of numerous downstream target genes that define adipocyte identity. During adipocyte maturation and differentiation, C/EBP-α and PPAR-γ are key adipogenic transcription factors. They promote adipogenesis by co-activating the expression of adipose-specific genes, such as AP2, and maintaining their expression at elevated levels [60].

In other studies, the expression of C/EBP-α was detected during the early stages of 3T3-L1 adipocyte differentiation and was associated with the maintenance of PPAR-γ expression [61]. The 3T3-L1 cell line is widely used as an adipocyte differentiation model to investigate the molecular mechanisms of adipogenesis [38]. In fact, C/EBP-α was present in both white and brown adipocytes, where its overexpression could induce adipocyte differentiation, and its levels are further elevated in the latter stages of differentiation [62]. The expression of C/EBP-α in fibroblasts induced adipogenesis, but only in the presence of PPAR-γ [63]. As a member of the nuclear receptor superfamily of ligand-activated transcription factors, PPAR-γ is crucial for adipocyte differentiation and fat formation [64][65]. It is selectively expressed in adipocytes, where it regulates their metabolism and fatty acid processing [66]. PPAR-γ increases the expression of C/EBP-δ and C/EBP-β during the early stages of differentiation and continues to elevate C/EBP-α levels to promote further differentiation [67]. As a transcriptional factor, PPAR-δ negatively regulated C/EBP-α and PPAR-γ promoters and positively regulated the heme oxygenase-1 (Ho-1) promoter [68].

Choi et al. demonstrated that, compared to other natural alkaloids, BBR exhibited strong inhibitory activity on adipogenesis in 3T3-L1 cells by down-regulating PPAR-γ and C/EBP-α expression in a concentration-dependent manner [38]. Meanwhile, Shou et al. found that BBR dose-dependently increased PPAR-δ levels, and silencing

PPAR-δ abolished BBR's effect on lipid accumulation [68]. After activating by BBR, the transcriptional functions of PPARδ were invoked. By activating PPAR-δ expression, BBR mediated the expression of downstream genes that were functionally involved in lipid accumulation [68].

### 3.1.2 Down-regulation of CREB and C/EBP-β transcriptional activity and Gal-3 activity by BBR

It has been demonstrated that the activation of cAMP-response element-binding protein (CREB) is crucial for initiating adipocyte differentiation process [69]. Activated CREB increases the expression of C/EBP-β in the early stages of adipogenic differentiation, triggering a cascade of transcriptional events that promote the activation of C/EBP-α and PPAR-γ2 genes, two crucial transcription factors involved in adipocyte differentiation [69]. In addition, 3-isobutyl-1-methylxanthine (IBMX) and cAMP analogues also can promote adipocyte differentiation [70].

Galectin-3 (Gal-3) is a member of the galectin family that binds carbohydrates or glycoproteins [71]. Gal-3 is expressed not only by macrophages but also by adipocytes, being high in the preadipocyte fraction and almost nil in differentiated adipocytes [72]. The mRNA and protein expression levels of Gal-3 increased in a time-dependent manner during the differentiation, proliferation and maturation of adipocytes, and Gal-3 is positively associated with obesity [73]. Moreover, it was found that recombinant human Gal-3 stimulates preadipocyte proliferation as well as DNA synthesis through lectin–carbohydrate interaction [72] Gal-3 deletion results in impaired terminal adipogenic differentiation, in the absence of obvious signs of adipocyte degeneration/death and changes in both the amount and distribution of body fat [74].

Florido et al. found that higher BMI is independently linked to elevated Gal-3 [75]. Many previous studies have shown that obesity significantly increases reactive oxygen species (ROS) levels in the vasculature, with NADPH Oxidase I (NOX1) being the most predominant source of vascular superoxide [76][77][78][79]. Padgett et al. indicated that Gal-3 is most highly expressed in the microvascular endothelium in obesity, and then confirm that Gal-3 is a novel mediator of NOX1-derived oxidative stress in the microvascular endothelium, Gal-3 oligomerization is required for NOX1 promotor activity [80]. Meanwhile, obesity-associated intermuscular adipose tissue (IMAT) is suggested to originate from PDGFRα+ mesenchymal progenitors [81]. Takada et al. suggest that obesity induces excess secretion of Gal-3 from immune cells upon muscle injury/regeneration, thereby activates the PPAR-γ signaling pathway, and then induces nuclear translocation and activation in PDGFRα+ cells, leading to a deviated differentiation process directed to ectopic intermuscular adipogenesis [82] [83].

Zhang et al. found that BBR primarily attenuated the IBMX and forskolin-induced phosphorylation of CREB, subsequently inhibiting the expression of C/EBP-β expression and the binding to cAMP responsive element (CRE), furthermore decreasing the expression of C/EBP-α and PPAR-γ2 during the early stage of 3T3-L1 preadipocyte differentiation. Additionally, the inhibition of the cAMP/PKA-mediated CREB signaling pathway is a key mechanism contributing to the anti-obesity effects of BBR [84].

Moreover, IMAT formation during muscle regeneration was significantly suppressed in Gal-3 knockout mice by reducing the expression of key factors involved in adipogenesis, such as PPAR-γ and C/EBP-α [83]. Wang et al. demonstrated that BBR inhibits the differentiation and proliferation of adipocytes and reduces adiposity by decreasing Gal-3 promoter activity as well as destabilizing Gal-3 mRNA. Furthermore, BBR up-regulated microRNA (miRNA) let-7d expression, which post-transcriptionally suppresses Gal-3 [85]. In addition, overexpression of Gal-3 induced a further up-regulation of PPAR-γ, Gal-3 is possibly a regulator of PPAR-γ and is upstream of PPAR-γ [85].

### 3.1.3 Influence the gene expression of adipokines related with metabolic regulation by BBR

Utami et al. showed that BBR inhibits complex I of the respiratory chain, contributing to its antioxidant capability by activating the translocation of nuclear factor erythroid

2–related factor 2 (Nrf2), which increases the gene expression of superoxide dismutase (SOD), glutathione peroxidase (GPX), and reduced glutathione (GSH) [86]. And then, BBR also have been shown to improve insulin sensitivity and stimulate glucose uptake through the activation of AMP-activated protein kinase (AMPK) [87] and the Toll-like receptor 4 (TLR4) signaling pathway [4][88]. All of these mechanisms suggest that BBR could influence the gene and protein expression of adipokines which are closely related to metabolic regulation. Meanwhile, Ye et al. considered that BBR can become a promising agent for treating metabolic syndrome and cardiovascular risk associated with obesity [89].

## 3.2 The activity of BBR in browning of white adipose tissue activation and metabolism regulation

Adipose tissue (AT) is primarily composed of white adipose tissue (WAT) and brown adipose tissue (BAT) with distinct physiological functions. In addition to WAT and BAT, beige adipose tissue, serves as a transitional form between WAT and BAT. WAT stores energy from glucose and fatty acids (FA) in the form of triglycerides within large unilocular lipid droplets in adipocytes and releases energy as free fatty acids (FFA) [90]. In contrast, BAT dissipates energy through non-shivering thermogenesis [91]. Beige adipocytes resemble BAT in appearance but most typically emerge from a Myf5-negative cell lineage, which is similar to that of WAT[92]. In fact, it is reported that beige adipocytes develop within subcutaneous WAT from a distinct subset of preadipocytes[93] or through the trans-differentiation of existing WAT[94][95], the latter is known as browning of WAT. BAT is predominantly located in the neck and subscapular regions, beige adipose tissue often found within AT, whereas WAT is distributed across various body parts [92].

The thermogenic capacity of BAT arises from its high mitochondrial content and elevated levels of uncoupling protein 1 (UCP1), which uncouples oxidative phosphorylation to generate heat instead of ATP[96]. The identification and analysis of UCP1 stimuli can greatly facilitate our understanding of AT thermogenesis, including the browning of WAT. Gong et al. summarize the stimuli that have activated UCP1 in recent decades, including physical and environmental factors (cold stimulation and exercise), factors related to traditional Chinese medicine (acupuncture, Chinese herbal formulas, Chinese medicinal herbs and their bioactive compounds), pharmacological agents, functional food and food ingredients, gut microbiota and pharmaceutical agents [97].

Browning of WAT refers to the transformation of WAT into beige adipocytes, which occurs under the influence of stimuli such as exercise, cold exposure and adrenergic receptor activation. The browning process is associated with the activation of various transcriptional factors including PRDM16, PPAR-γ, PGC-1α, COX2, PG, FGF21, BMP7, MANF, activin E, C/EBP-β, myokines, zinc finger protein 516, ATGL, adiponectin, and irisin. More recently, there have been a number of findings about the browning of WAT. Kim et al. suggested that beta-hydroxybutyrate (β-HB) reduced intracellular lipid accumulation by activating mitochondrial biogenesis, enhancing lipolysis and stimulating the expression of thermogenic and fat browning genes [98]. Wang et al. unravel that EDN3/EDNRB signaling induces the thermogenic differentiation of white adipose progenitor cells via activating intracellular cAMP and EPAC1-ERK-mediated pathways [99]. In human visceral adipocytes, Neira et al. investigated that Fibronectin type III domain-containing protein 5 (FNDC5) /irisin triggered mitochondrial biogenesis (TFAM) and fusion (MFN1, MFN2, and OPA1) while inhibiting peripheral fission (DNM1L and FIS1) and mitophagy (PINK1 and PRKN), and then stimulating the browning of WAT process in response to cold and obesity [100]. Wang et al. found that the β3-adrenergic receptors (β3ARs)-activated mTOR-lipin1 axis mediates browning of WAT [101]. Zhou et al. demonstrated that the nuclear receptor subfamily 2 group F member 6 (NR2F6) played a pivotal role in brown adipogenesis and energy homeostasis, NR2F6 transcriptionally regulated PPAR-γ expression to promote adipogenic process in BAT [102].

The expression of BAT marker genes (PGC-1α and UCP1), transcription factors (PPAR-α, nuclear respiratory factor 1 and mTFA) and mitochondrial biogenesis (ATPsyn, COXIV and Cyto C) were concurrently activated in fractionated and differentiated primary IWAT which were treated with BBR [103]. Meanwhile, BBR chloride, a dual topoisomerase I and II inhibitor，Ferdous et al. found that it significantly up-regulated UCP1 gene expression and down-regulated ATP production in BAT which reduced adipocyte content by initiating thermogenesis [104]. The activation of AMPK and PGC-1α also plays a significant role in the thermogenic induction by BBR [103]. Yao et al. indicated that BBR attenuates the abnormal ectopic lipid deposition in skeletal muscle by promoting the mitochondrial biogenesis and improving fatty acid oxidation in an AMPK/PGC-1α dependent manner[105]. Ling et al. demonstrated that BBR epigenetically acts through the AMPK-α-ketoglutarate (α-KG)-PRDM16 axis to promote brown adipogenesis and BAT thermogenesis. BBR increases the transcription of PRDM16 by inhibiting DNA demethylation of the PRDM16 promoter, likely driven by AMPK activation and the production of the tricarboxylic acid cycle intermediate α-KG[106]. Sun et al. provided both in vivo and in vitro evidence that BBR promotes the browning of WAT to enhance energy expenditure and weight loss by stimulating $NAD^+$-dependent deacetylase SIRT1 activity and inducing autophagy in an autophagy protein 5-dependent manner, as well as inducing the production and secretion of FGF21 [107]. Growth differentiation factor 15 (GDF15), a distant member of the transforming growth factor-β superfamily, forms a ternary complex with glial cell-derived neurotrophic factor family receptor alpha-like (GFRAL) and the receptor tyrosine kinase in the hypothalamus to reduce energy intake by suppressing appetite. Peripheral GDF15 can also promote thermogenesis and lipolysis while reducing adipose tissue mass [108]. Li et al. found that BBR increases serum levels of GDF15 and have shown that BBR lowers body weight by up-regulating GDF15 secretion and expression via the activation of the integrated stress response (ISR) in BAT [108].

Above mentioned evidences consistently identified that BBR and its derivatives are crucial factors in promoting adaptive thermogenesis by activating BAT activity and the browning of WAT process.

**3.3 The hypolipidemic activity of BBR**

Obesity alters lipids metabolism due to high triglycerides, high LDL-C, and high total cholesterol, leading to endothelial dysfunction and atherosclerosis, contribute to atherosclerotic diseases [109] [110] [111]. Foam cell formation is the initial pathological process of atherosclerotic diseases. Guan et al. found that BBR activates the AMPK-SIRT1-PPAR-γ pathway, inhibiting the expression of lectin-like oxidized LDL receptor 1 (LOX-1), which impedes foam cell formation [112]. Foam cell accumulation leads to fibrous plaques that evolve into atherosclerotic plaques [113]. Through activing nuclear factor erythroid 2 - related factor 2 (NRF2)/recombinant solute carrier family 7 member 11 (SLC7A11)/glutathione peroxidase 4 (GPX4) pathway and inhibiting ferroptosis in endothelial cells acting as an ACSL4 inhibitor, BBR can resist oxidative stress, inhibit ferroptosis, reduce plaque area and stabilize plaque [114] [115].

Meanwhile, obese patients mainly show expansion of adipose tissue and hypertrophy of adipocytes [116]. Adipocyte hypertrophy leads to the synthesis of adipokines mainly including visfatin (Vis), resistin (Res), apelin (Ap) and omentin (Ome) that have different physiological functions [117]. Recent Bandala's research has shown that BBR lowers the harmful adipokines expression of Vis and Ap, and enhanced the protective adipokines of Ome, which can be explained by BBR enhances lipid metabolism and fatty acid oxidation, and reduces adipogenesis and IR in adipocytes, thereby alleviating obesity and inflammation [56].

Two clinical trials in patients with dyslipidemia have demonstrated that BBR can reduce triglycerides by 35% and 22%, serum cholesterol by 29% and 16%, and LDL-C by 25% and 20%, respectively [118] [119]. Du et al. certified that BBR reduced triglyceride level and lipid droplet volume in hypertrophic adipocytes, further demonstrated that BBR inhibited miR-27a levels which induced IR status in serum from obese mice [54].

Studies have also linked the hypolipidemic activity of BBR to low density lipoprotein receptor (LDLR) mediated cholesterol metabolism. BBR can up-regulate LDLR expression in liver cells through AMPK-dependent Raf-1 activation [120]. The BBR can also induce the stabilization of LDLR mRNA by activating the extracellular signal-regulated kinase (ERK) signaling pathway and down-regulating the mRNA decay-promoting factor heterogeneous nuclear ribonucleoprotein D (hnRNP D) [121] [122]. Lee et al. showed that the effect of BBR on LDLR was blocked by a c-Jun N-terminal kinase (JNK) inhibitor, suggesting that JNK activation was involved [123]. Similarly, BBR (at 50 and 100 mg/kg) in a tilapia model reduced plasma lipid levels but increased the expression of the PPAR-α and carnitine palmitoyltransferase 1 (CPT-1) genes in the liver, leading to reduced lipid accumulation [124].

Additionally, BBR treatment can increase the conversion of cholesterol to bile acid, thereby reducing blood cholesterol levels [125]. CYP7A1 and SREBP are considered the classical rate-limiting enzyme and transcriptional factor for the conversion of cholesterol to bile acid [126], and the hypolipidemic effect of BBR may be related to its up-regulation of SREBP2 and CYP7A1 expression and the promotion of bile acid metabolism [127].

Previous studies have shown that lacteals are responsible for draining chylomicrons, which are triglyceride-rich and processed by enterocytes, into the lymphatic system via the interstitial space among the lymphatic endothelial cells (LECs) on the lacteals [128]. Zhang et al. demonstrated that diet-induced obesity can be inhibited by converting the discontinuous and open "button-like" LEC junctions into continuous and closed "zipper-like" junctions, thereby reducing the entry of chylomicrons into the lacteals [129]. Based on these findings, Wang et al. provided novel evidence for one of the hypolipidemic mechanism of BBR. They determined that BBR inhibits lipid absorption by promoting lacteal junction zippering and verified that BBR promoted the formation of mature LEC junctions via the suppression of the Ras homolog gene family member A (RhoA)/Rho-associated kinase 1 (ROCK1) signaling pathway [53].

### 3.4 Regulation of the gastrointestinal microbiota (GM) by BBR

Previous studies had revealed that BBR have antibacterial activity through the inhibition of protein and DNA synthesis pivotal to bacterial growth and proliferation. Consequently, it effectively halts the division and development of various common bacterial strains [130] [131] [132]. Ye et al. confirmed that BBR exhibits antimicrobial effects on hemolytic Streptococcus, Staphylococcus aureus, Neisseria gonorrhoeae, and Freund's Shigella, and can improve leukocyte phagocytosis [133]. And then, BBR can significantly reduce the abundance of Proteobacteria such as Desulfovibrio and Enterobacter cloacae, inhibiting lipopolysaccharide (LPS) production and effectively attenuating serum LBP, a biomarker of circulating LPS, thereby improving metabolic endotoxemia [134]. Disturbances in the Firmicutes/Bacteroidetes ratio in the gut microbiota have been associated with various diseases, including obesity[135]. Xie et al. showed that administering BBR at a dose of 200 mg/kg for six weeks significantly reduced the relative abundances of the phyla Bacteroidetes and Firmicutes in the gut of HFD-fed mice, and systematically increased the expression of fasting-induced adipose factor genes in visceral AT (VAT) [136]. BBR has also been shown to enrich the population of butyrate-producing bacteria in the gut microbiota, promoting butyrate synthesis via the acetyl-CoA-butyryl-CoA-butyrate pathway. Subsequently, butyrate enters the bloodstream, reducing lipid and glucose levels [134]. Wu et al. demonstrated that BBR and its derivatives exert lipid-lowering effects by mediating gut microbiota, identifying Blautia as a critical commensal genus for its anti-hypercholesterolemic action. They also noted that baseline levels of Alistipes and Blautia could precisely predict the efficacy of BBR against hypercholesterolemia[137]. Sun et al. reported that BBR increased the Bacteroidetes/Firmicutes ratio and the proportion of SCFA-producing bacteria by regulating the GM-gut-brain axis, which in turn promoted the expression of GLP-1 in intestinal L cells [138]. Yang et al. investigated the effects of BBR metabolites on GLP-1 secretion, demonstrating that BBR significantly increased the production and glucose-stimulated secretion of GLP-1 in GLUTag cells by alleviating cell death, oxidative stress, mitochondrial dysfunction, and reversing inflammation-induced inhibition of the Akt signaling pathway [139]. Sun et al. also revealed that BBR can bind to bitter-taste receptors (TAS2Rs) to promote GLP-1 secretion and enhance downstream Gα-gustducin/Gβ1γ 13 signaling in tuft cells in the gut. This activity also increases IL-25 production and repairs the compromised gut barrier integrity induced by obesity [52].

Similarly, Wang et al. found that BBR strengthens the intestinal barrier by ameliorating endoplasmic reticulum (ER) stress and reducing goblet cell apoptosis through decreased mucin-2 expression [140]. Akkermansia muciniphila (A. muciniphila) has been identified as a mucin-degrading bacterium in the mucus layer [141]. Previous studies have shown a reduced abundance of A. muciniphila in obese humans [142]. Wang et al. demonstrated that BBR can significantly increase the abundance of bacteria of the phylum Verrucomicrobia, particularly A. muciniphila, which leads to the regulation of tight junction proteins and protection of intestinal barrier integrity [140]. Moreover, Chen et al. showed that administration of BBR could alleviate intestinal barrier dysfunction of glucolipid metabolism disorders (GLMDs) by increasing the number of colonic glands and goblet cell mucus secretion, promoting the proliferation of beneficial microbiota, and altering the levels of tryptophan metabolites, intestinal tight junction proteins and intestinal immune factors [143].

Additionally, BBR acts as an α-glucosidase inhibitor, breaking down carbohydrates into monosaccharides and diminishing the absorption of dietary carbohydrates. BBR's α-glucosidase inhibitory activity also confers significant antimicrobial effects by inhibiting the assembly function of FtsZ and halting bacterial cell division in the gastrointestinal tract [144].

### 3.5 BBR in adipose tissue macrophages (ATMs) recruitment and polarization

Macrophages are typically classified into M0, M1 and M2 types: M0 macrophages are quiescent cells capable of phagocytizing cell debris, M1 macrophages exhibit phagocytic and pro-inflammatory properties and M2 macrophages are associated with attenuated inflammation and tissue deposition [145]. Numerically the population of ATMs expands from 10% of all cells in lean AT to more than 50% in severe obesity in mice [146] [147]. Macrophage numbers increase in AT in obese humans, with the percentage of ATMs rising from 5%–10% in the lean healthy status to 40%–50% of the stromal cells in the obese setting[148]. Russo et al. give a conclusion that ATMs are key mediators of meta-inflammation, IR and impairment of adipocyte function with progressive obesity, which consistent with above-mentioned clinical data [149]. Based on this, many groups have studied the relationship and mechanism between BBR and ATM.

Noh et al. reported that BBR significantly reduced body weight, adipocyte size, fat deposition in the liver, and ATM infiltration in C58BL/6 mice fed with high fat diet (HFD) [150]. In addition, an increased number of the CD206$^+$ M2 ATM cells were noticed. It was proposed that BBR led to weight loss in the HFD model by modulating ATM recruitment and polarization via chemotaxis inhibition [150]. Gong et al. had a study unveiled that BBR decreased monocyte chemoattractant protein-1 (MCP-1) expression, while diminishing macrophage infiltration and decreasing M1 proportion as determined by RT-qPCR [151]. Additionally, Wu et al. camouflaged BBR-loaded PLGA nanoparticles (BBR NPs) with M2 macrophage cell membrane and loaded them with Man, which induce the polarization of M1 macrophages toward the M2 phenotype upon reaching the plaque, thereby mitigating local inflammation and exerting anti-inflammatory effects, increasing collagen levels in the aorta, reducing lipid deposition in plaques and improving plaque stability [152]. Lin et al. also verified that BBR treatment in mice fed with the high-fat diet increased energy metabolism, glucose tolerance, and expression of UCP1, and reduced expression of pro-inflammatory cytokines, macrophage recruitment, and resulted in M2 macrophage polarization in WAT [153]. These findings underscore the pivotal role of macrophages in the development of obesity, rendering their metabolic modulation a promising therapeutic target [154].

Zhang et al. identified that BBR could reduce the acetylation of p65 at site Lys310, thereby weakening the translocation of p65 and inhibiting p300/ac-p65Lys310 signaling in macrophages and leading to the decline of NF-κB activity [155]. NLRP3 (Nod-like receptor family pyrin domain containing 3) inflammasome has been reported to contribute to obesity-induced inflammation and insulin resistance. Zhou et al. demonstrated that BBR significantly inhibited the

NLRP3 inflammasome activation and interleukin-1β (IL-1β) release triggered by saturated fatty acid palmitate (PA) in macrophages. This inhibitory effect was mediated through the induction of autophagy in ATMs in an AMPK-dependent manner, highlighting the potential of BBR in modulating inflammatory responses[156]. Zhang et al. presented a novel perspective suggesting that BBR, when combined synergistically with isoliquiritigenin (ISL), upregulates the IRS1-PI3K-Akt insulin signaling pathway, thereby enhancing glucose uptake and mitigating the interaction between macrophages and adipocytes. This combination also reduces the accumulation and infiltration of M1-ATMs[157].

## 3.6 The activity of BBR in improving adipose tissue inflammation

Chronic inflammation may contribute to a variety of disorders, such as diabetes, obesity, rheumatoid arthritis, atherosclerosis and cancer[158][159]. WAT accumulation and inflammation contribute to obesity by inducing insulin resistance[157]. In the obese state, long-term overactivation of ATMs results in chronic low-grade inflammation and a shift in the adipose immune landscape[160]. The classically activated ATMs (M1-ATMs) secrete a variety of cytokines including IL-6 and TNF-α, and chemokines, such as macrophage inflammatory protein-1α (MIP-1α) and MCP-1, which result in low-grade chronic inflammation[149].

Li et al. demonstrating that BBR mitigates adipose tissue inflammation by suppressing macrophage infiltration into eWAT in HFD-fed mice[160]. Similarly, BBR significantly down-regulated the expression of the NLRP3 inflammasome and its associated molecules in macrophages, thereby mitigating NLRP3 inflammasome activation-induced macrophage M1 polarization and inflammation[161]. In addition, recent research has shown that BBR increased the gene expression of the antioxidant components erythroid 2-related factor 2 (NRF-2), heme oxygenase 1 (HO-1), and glutathione-S-transferase-α (GST-α), and reduced inflammation by decreasing Toll-like receptor-2 (TLR-2), myeloid differentiation protein-88 (MYD-88), interleukin-1β (IL-1β), tumor necrosis factor-α (TNF-α), and interleukin-8 (IL-8) gene expression[124].

The deacetylase SIRT3 regulates mitochondrial biogenesis and function, and its expression level and activity are decreased in obesity[162]. Notably, BBR suppresses TNF-α-mediated inflammation in differentiated adipocytes, at least partially by directly binding and activating the deacetylase SIRT3 and suppressing the activation of the MAPK and NF-κB signaling pathways[160]. Meanwhile, Meng et al. firstly identified IRGM1 as the direct anti-inflammatory target of BBR in cell culture-activity-based protein profiling (SILAC-ABPP). Further clarification that the molecular mechanism of BBR's anti-inflammatory activity was inhibiting the PI3K/AKT/mTOR pathway by targeting IRGM1[163]. Yang's results revealed that berberrubine (BBB) and palmatine (PMT), out of BBR metabolites, alleviated cell death, oxidative stress, and mitochondrial dysfunction and effectively reverse inflammation-induced inhibition of the Akt signaling pathway in mice with obesity induced by HFD[139].

Obesity and IR are common characteristics of polycystic ovary syndrome (PCOS)[164]. As one of the common causes of obesity in women, PCOS also associated with AT inflammatory response and apoptosis[165]. Shen et al. had some findings that BBR may alleviate the inflammatory response of PCOS rats by regulating the expression of inflammation-related genes, including TLR4, LYN, NF-κB, TNF-α, IL-1 and IL-6. Additionally, BBR may improve PCOS via the PI3K/Akt/NF-κB signaling pathway[166].

## 4. Better formulation to improve the anti-obesity activity of BBR

Despite the beneficial effects and high safety profile of BBR, its poor bioavailability due to low intestinal absorption remains a significant limitation in its clinical application. BBR will need improved formulations or completed targeted delivery systems to enhance its potency before it can effectively compete with GLP-1 agonists in weight loss clinical trials. In addition, new salt or new crystal form of BBR may also improve its bioavailability.

### 4.1 Rational combinations of active compounds from herbs

For compounds with poor absorption and low bioavailability, such as BBR, combining active ingredients from herbs offers a potential approach to enhance therapeutic effects, allowing BBR to be effective at much lower doses.

Insulin - sensitizing phytonutrients, such as cinnamaldehyde and curcumin, are known inhibitors of protein serine/threonine phosphatases and protein tyrosine phosphatases. These inhibitors increase phosphorylation of Akt2 at T450 and Y475, priming Akt2 for subsequent activation via insulin-stimulated phosphorylation at S474 [167]. Urasaki et al. presented a rational composition of phytonutrients (F2) comprising cinnamaldehyde, curcumin, and BBR. This composition protects against obesity and pre-diabetic conditions in a diet-induced obesity murine model, providing an effective means to improve insulin sensitivity without increasing adiposity [168]. Similarly, Yue et al. demonstrated that astragalus polysaccharides (APS) combined with BBR were more effective than APS or BBR alone in reducing obesity and modulating the gut microbiota in HFD-fed mice [169]. Meanwhile, BBR acted synergistically with ISL to significantly ameliorate obesity and dyslipidemia in DIO mice, and can be a promising and effective strategy for improving obesity-induced adipose inflammation and insulin resistance [157]. Additionally, dual and triple combinations of BBR, catechin, and capsaicin，have shown better potential than single active ingredients[170]. In obese and diabetic patients, BBR combined with silymarin significantly reduced fasting blood glucose and insulin, the IR index, HDL and LDL, triglycerides, uric acid, BMI, and the WHR compared with the use of BBR alone [171].

In Lin's study, a water-soluble tetraanionic macrocycle has been revealed to be highly biocompatible and efficiently included encapsulate discrete electron-deficient aromatic compounds (including BBR and palmatine) through parallel-arrangement encapsulation, and this macrocycle is able to inhibit the bitter taste of the two drugs [172].

## 4.2 A co-crystalized BBR-ibuprofen

The drug co-crystal refers to a crystal composed of the active pharmaceutical ingredient (API) and a co-crystal former (CCF) bonded by non-covalent interactions

[173]. Drug co-crystals can enhance the physicochemical and biopharmaceutical properties of APIs [174].

Wang et al. prepared a 1:1 co-crystal of BBR and ibuprofen (termed BJ) using drug salt metathesis and co-crystal technology. Compared to the common BBR salt (BBR chloride dihydrate, BCl·2H2O), BJ demonstrated a threefold increase in bioavailability in vivo and was more effective in treating obesity and related metabolic disorders in the db/db mice model [175]. It was reported that BJ inhibited the expression of IKKε and TBK1 in the epididymal WAT (eWAT) of db/db mice, but had no effect on these proteins in BAT. TBK1 directly inhibits the phosphorylation of AMPK-α at the Thr172 site (p-AMPK-α) in adipocytes, regulating UCP-1 and mitochondrial biogenesis [176]. IKKε reduces cAMP levels in adipocytes and inhibits cAMP-mediated β-adrenergic signal transduction, leading to catecholamine resistance and reduced lipolysis and thermogenesis [177]. Thus, studies with BJ revealed that BBR could promote mitochondrial biogenesis and increases adipocyte sensitivity to catecholamines by inhibiting the activity of non-canonical IκB kinases TBK1 and IKKε, and induces AMPK phosphorylation [175].

## 4.3 Delivery systems

Numerous researches consistently suggest that bomimetic nanomaterials by fusing natural cell membranes onto nanoparticles (NPs) [178] that coat cell membranes have been extensively researched in recent years due to it enables nanoparticles to evade the immune system [179][180], extend their circulation in the bloodstream [181][182] and facilitate the targeted uptake by specific homologous cells [183]. Based on this, Wu et al. developed BBR NPs@Mannose (Man)/M2 nanoparticles to enhance the bioavailability of BBR [152]. Meanwhile, Xu et al. develop a novel nanoemulsion (NE) that increase the oral bioavailability of BBR in rats by 212.02%, BBR-loaded NE protect BBR against the intestinal metabolism mediated by CYP2D6 and CYP3A4 [184].

Additionally, Mirhadi et al. describes different types of nanocarriers (polymeric based, magnetic mesoporous silica based, lipid based, dendrimer based, graphene based, silver and gold nanoparticles) have been used for encapsulation of BBR to improve its bioavailability [185]. Erythrocytes have been extensively studied for their biocompatibility, safety, the preferential uptake and high drug loading efficiency in vascular carriers [186][187]. An erythrocyte-based drug delivery system for BBR has been developed to target macrophages and increase biocompatibility, enhance therapeutic efficacy in combating inflammation and hypolipidemic effect, achieve long circulation and sustain release of BBR [188].

**4.4 Derivatives of BBR**

Tetrahydroberberrubine (THBru, figure 2), a derivative of BBR, was found to markedly ameliorate obesity, dyslipidemia, and decrease Lee's index, fat mass in eWAT and BAT in HFD-induced obese mouse model through the activation of PGC1α-mediated thermogenesis. Notably, THBru was demonstrated to exhibit superior efficacy compared to BBR at the same dose [189].

**Figure 2 Chemical Construction of THBru**

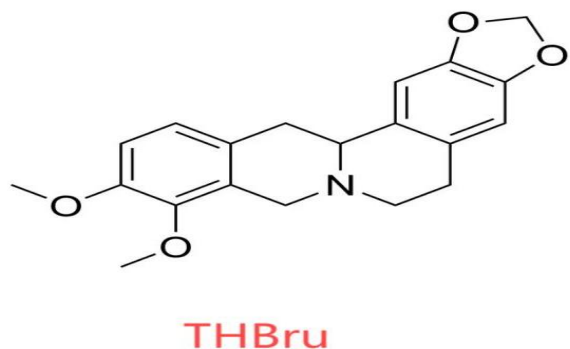

Oxyberberine (OBB, figure 3), a gut microbiota-mediated oxidative metabolite of BBR. Li et al. firstly identified that OBB remarkably and dose-dependently attenuated the clinical manifestations in HFD-induced obese NAFLD rats, which was superior to BBR of the same dose (100 mg/kg). Meanwhile, compared with BBR, they further proof that OBB treatment remarkably up-regulates UCP-1 protein expression to increase expenditure of energy by exhibiting superior hyperphosphorylation of

AMPK in vivo, alleviated inflammation and IR to maintain lipid homeostasis by significantly inhibiting aberrant phosphorylation of IRS-1, down-regulating the mRNA expression of MCP-1, Cd68, Nos2, Cd11c, while up-regulating the downstream protein expression and phosphorylation (PI3K, p-Akt/Akt and p-GSK-3β/GSK-3β) [190].

**Figure 3 Chemical Construction of OBB**

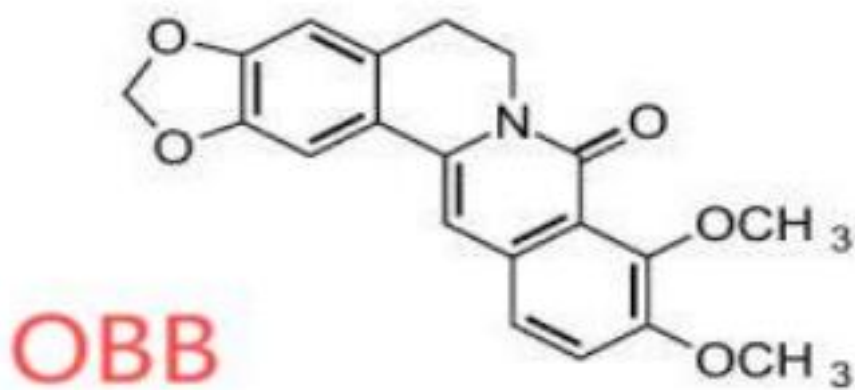

9-N-alkyltetrahydroberberine (figure 4) is also a derivatives of BBR. Just administration of this substance at a dose of 15 mg/kg for four weeks significantly improved the insulin sensitivity of obesity mice, and the mass of AT was demonstrated lower than in untreated obesity mice [191].

**Figure 4 Chemical Construction of 9-N-alkyltetrahydroberberine**

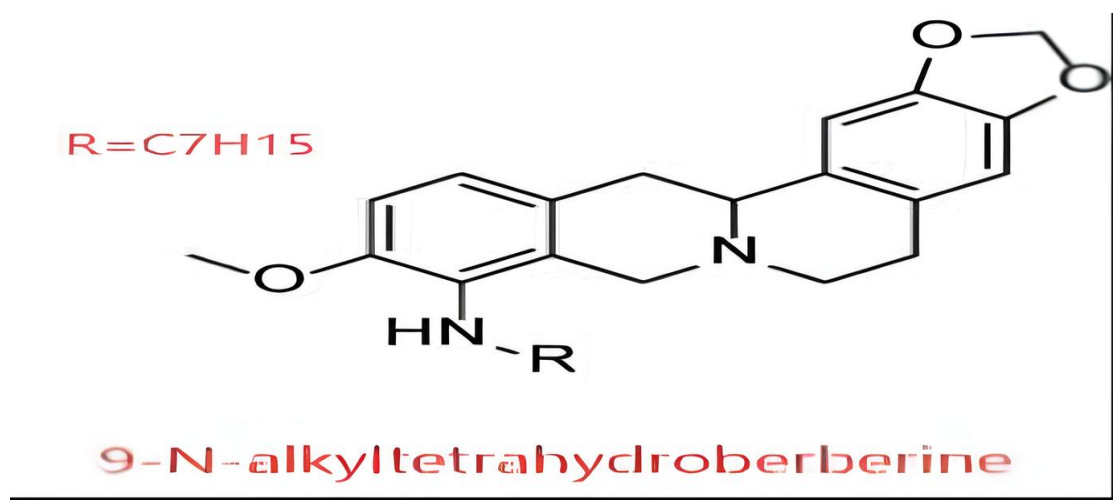

Similarly, a new derivative of BBR, 9-(hexylamino)-2,3-methylenedioxy-10-methoxyprotoberberine chloride (SHE-196, figure 5) decreased total body weight and interscapular fat mass, and increased interscapular brown fat activity in obese T2DM mice[192].

**Figure 5 Chemical Construction of SHE-196**

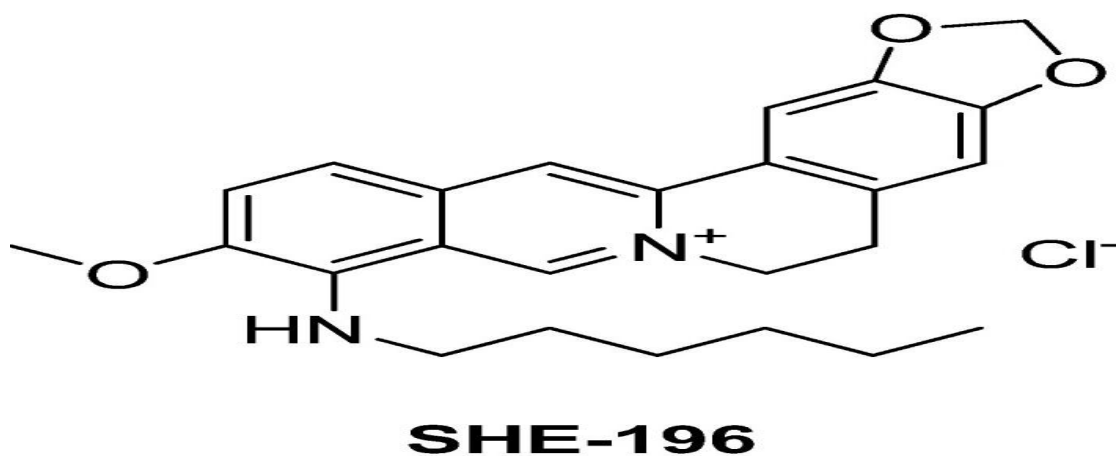

In addition, the reduced derivative of BBR, dihydroberberine (dhBBR, figure 6), has been suggested as a natural alternative of BBR to improve bioavailability[193]. A randomized, double-blind, placebo-controlled investigation determined that four doses of a 100 mg dose of dihydroberberine and 200 mg dose of dihydroberberine produce significantly greater concentrations of plasma BBR across of two-hour measurement window when compared to a 500 mg dose of BBR or a placebo (BBR level: PLA: 0.22 ± 0.18 ng/mL, B500: 0.4 ± 0.17 ng/mL, D100: 3.76 ± 1.4 ng/mL, D200 : 12.0 ± 10.1 ng/mL; BBR AUC: PLA: 20.2 ± 16.2 ng/mL × 120 min, B500: 42.3 ± 17.6 ng/mL × 120 min, D100: 284.4 ± 115.9 ng/mL × 120 min, D200: 929 ± 694 ng/mL)[194].

**Figure 6 Chemical Construction of dhBBR**

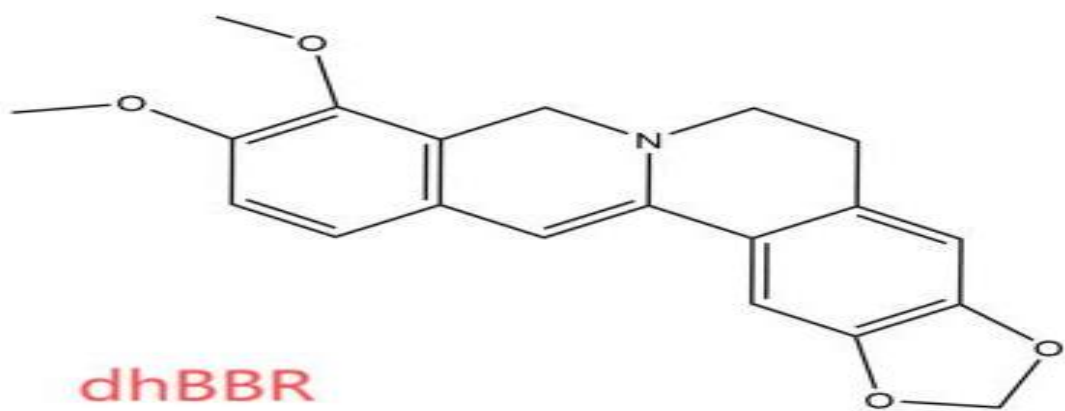

dhBBR

The bioavailability, however, could be significantly enhanced through introduction of lipophilic substituents. Among all lipophilicity-enhancing strategies, C9-O-Me substitution is the most intensively studied one, which is primarily due to its synthetic convenience. Teng et al. herein reported the first synthetic study on 9-O-arylated BBR (figure 7) via copper-catalyzed CAr–O coupling reactions, and provided a new perspective on the preparation of more diverse BBR derivatives in the future [195].

**Figure 7 Chemical Construction of 9-O-R BBR**

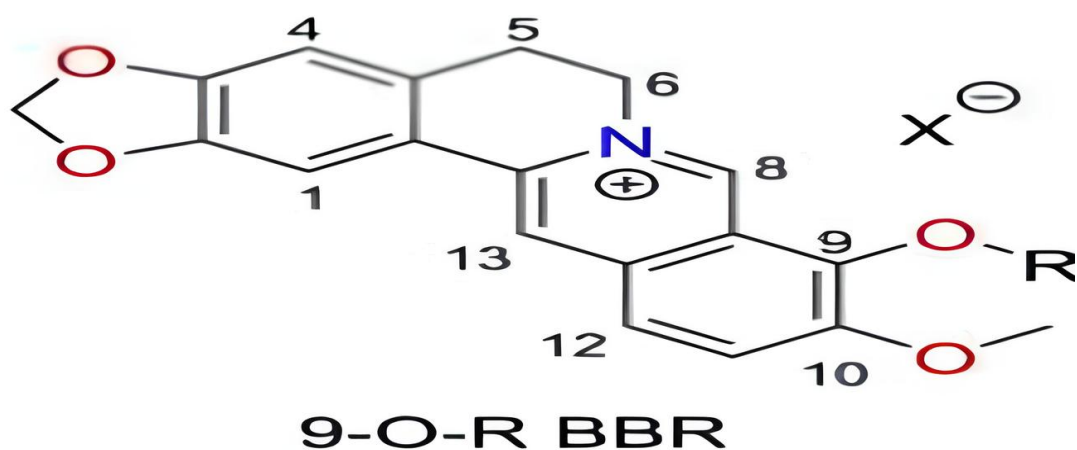

9-O-R BBR

Cheng et al. synthesized the 8,8-dimethyldihydroberberine (Di-Me, figure 8) with improved stability, and bioavailability over dhBBR. Administered to db/db mice with a dosage of 50mg/kg, Di-Me effectively reduced random fed and fasting blood glucose, improved glucose tolerance, alleviated insulin resistance and reduced plasma triglycerides, with better efficacy than dhBBR at the same dosage [196].

**Figure 8 Chemical Construction of Di-Me**

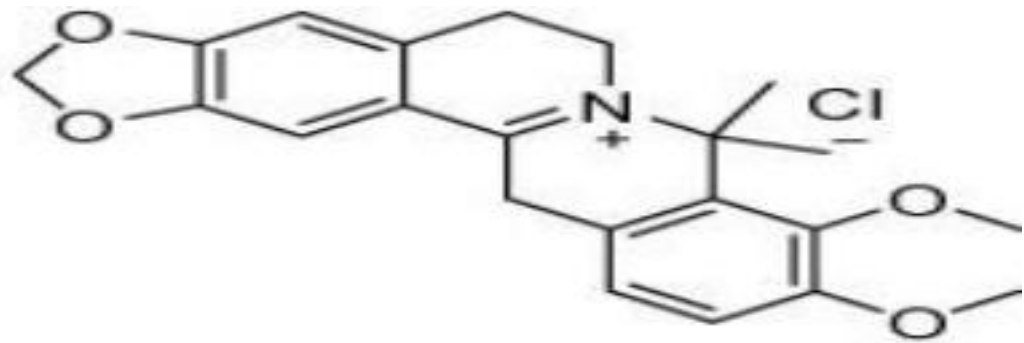

Di-Me

## 5 Summary

BBR, a quaternary isoquinoline alkaloid with potent pharmacological properties, is currently attracting significant attention for obesity treatment. Numerous studies have confirmed BBR's potential to mitigate obesity through various mechanisms (Figure 9). One key mechanism involves down-regulating genes that promote adipocyte proliferation and differentiation. Meanwhile, BBR influence other gene expression of adipokines related with metabolic regulation. Furthermore, BBR regulates the activation of BAT and the process of browning of WAT and metabolism, contributing to obesity treatment. And then, BBR regulates the lipid and plaque metabolism. In addition, BBR can alter the composition of gastrointestinal microbiota and strengthen intestinal barrier. Eventually, BBR promotes the polarization of ATMs towards the M2 phenotype, suppresses the activation of inflammatory signaling pathway and inflammatory cytokines.

Recent breakthroughs in researches include BBR's regulation of AMPK and PGC-1α activation, activation of the AMPK-α-KG-PRDM16 axis, stimulation of SIRT1 deacetylation activities, induction of FGF21 secretion through autophagy, and increased serum levels of GDF15. Additionally, BBR can alleviate AT growth by inducing enzymes that enhance glucose and FA uptake. Recent advancements also show that BBR promotes dyslipidemia metabolism by up-regulating LDLR expression via JNK activation, down-regulating LOX-1 expression through the AMPK-SIRT1-PPAR-γ pathway, and up-regulating expression of SREBP2 and CYP7A1. BBR has also been shown to promote LEC mature junction formation via suppression of the RhoA/ROCK1 signaling pathway. Studies indicate that BBR may improve IR and address obesity-associated cardiovascular diseases by influencing gut microbiota composition. Notable findings include BBR's ability to increase GLP-1 expression via TAS2Rs signaling pathway.

Despite its promise, BBR's low bioavailability and poor oral absorption have limited its clinical application. Recent research on the anti-obesity effects of BBR highlights rational combinations of herbs based on functional complementarity, co-crystal BBR-ibuprofen compounds, the construction of drug delivery systems targeting WAT and the development of derivatives (Figure 9). These advancements provide a theoretical basis for the more precise clinical use of BBR in promoting weight loss.

BBR has significant potential as a new drug for treating metabolic syndrome and hereditary endocrine diseases. However, this review only covers a portion of the recently discovered mechanisms of BBR against obesity. Most experiments have been conducted in rodents, whose fat distribution and metabolism differ from those of humans. Furthermore, the specific regulatory mechanisms of fat activation are complex and require further research to elucidate their benefits for human health. Although weight loss has been observed in patients enrolled in some clinical studies involving the common BBR chloride salt, these trials are not specifically designed for obesity treatment. In the future, more high-quality clinical trials need to be conducted to investigate the dose-response effects, optimize the optimal treatment period, and uncover the molecular mechanisms of BBR. This review can provide reference for further research and rational use of BBR in the treatment of obesity.

**Figure 9 Summary of BBR**

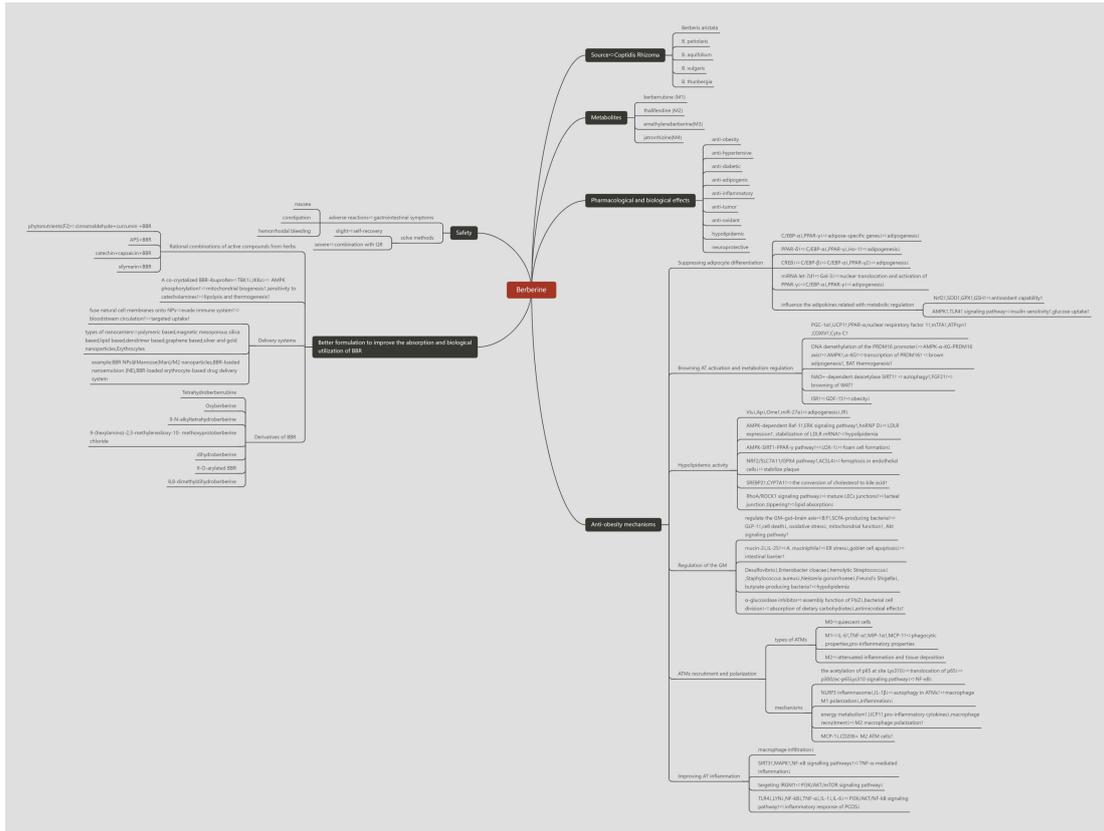

## 6. Declarations

(1) Ethics approval and consent to participate

Not applicable.

(2) Consent for publication

The review in this paper have not been published previously in whole or part,all authors unanimously consent for publication.

(2) Availability of data and materials

Not applicable.

(3) Competing interests

The authors declare that they have no known competing financial interests or personal relationships that could have appeared to influence the work reported in this paper.

(4) Funding

This work was supported by Priority Union Foundation of Yunnan Provincial Science and Technology Department and Kunming Medical University (No.202201AY070001-066), Funding of the "famous doctors" project of the support

(6)Authors' contributions

Y.K. and H.K.Y. were responsible for manuscript writing. R.N. and F.Z. were responsible for searching literature. X.Y.Z. and H.T.Z. provided manuscript writing guidance for the review. X.N. provided financial support. All authors have read and approved the final manuscript.

(7) Acknowledgements

This work was supported by Priority Union Foundation of Yunnan Provincial Science and Technology Department and Kunming Medical University (No.202201AY070001-066), Funding of the "famous doctors" project of the support plan for the talents of Xingdian (No.RLMY20220005), The scientific and technological innovation team of Kunming Medical University （CXTD202209）,Yunnan Provincial Endocrinology and Metabolism Clinical Medicine Center (YWLCYXZXXYS20221005), Clinical Collaboration Project of Traditional Chinese and Western Medicine for Major Difficult Diseases of Yunnan Province (300073).